\documentclass[a4paper]{jpconf}
\usepackage{graphicx}
\begin{document}
\title{Kinematics of AGN Jets}

\author{Eduardo Ros 
}

\address{
Max-Planck-Institut f\"ur Radioastronomie, Auf dem H\"ugel 69, D-53121 Bonn, Germany}

\ead{ros@mpifr.de } 

\begin{abstract}
The fine-scale structure and the kinematics of relativistic active galactic nuclei (AGN) jets have been studied by very-long-baseline interferometry at
very high resolutions since 1998 at 2\,cm wavelength for a 
sample of over a hundred radio sources (VLBA~2cm~Survey and MOJAVE programs). 
Since 2007, this is being complemented by the TANAMI project, 
based on southern 
observations with the Australian LBA at 3.6\,cm and 1.1\,cm wavelengths. 
From our observation campaign, we find that most of the radio jets 
show linear morphologies at parsec-scales, but some of
show curvature and non-radial motions.  Features are 
observed to move at highly relativistic speeds, with Lorentz
factors extending above values of 30.
We also provide a brief description of the relationship of our 
radio findings with the AGN observations by the
new \textit{Fermi Gamma-ray Space Telescope}.
\end{abstract}

\section{Introduction}

Through the combination of radio telescopes spread world wide, 
Very-Long-Baseline Interferometry (VLBI) yields the highest resolutions 
and the most accurate positions in astronomy.  At a wavelength of 2\,cm, 
i.e., the Very Long Baseline Array (VLBA) provides 
resolutions of the order of 1~milliarcsecond (mas) (that is, of parsec 
scales for quasars and BL\,Lac objects at moderate redshifts, and far 
below this for the closer radio galaxies).  VLBI has a large sensitivity 
when using large telescopes, but is in general limited to high brightness 
temperature sources (above $10^6$\,K) and to narrow fields (not larger than 
the arc second, generally limited by the correlator parameters rather than 
by the beam size of the single telescopes involved in the observations), 
making the source structure to be easily resolved out. It is, thus, 
an invaluable 
technique to study the highly relativistic jets AGN.

Due to relativistic effects, the intrinsic quantities in the AGN have to 
be deduced from the observed ones, which are affected by Doppler boosting 
and light travel time effects.  Table \ref{table:jetphys} provides the basic 
relationships between the measured and the observed quantities.  As a 
result of Doppler boosting, most of the AGN jets appear one-sided. Some 
close galaxies with small Doppler factors 
show jet and counter-jet with more symmetric features.

\begin{table}
\caption{\label{table:jetphys} Basic jet properties and radio/VLBI measurements}
\begin{center}
\[
\resizebox{\textwidth}{!}{%
\begin{tabular}{lr@{\,=\,}llr@{\,=\,}l}
\br
\multicolumn{3}{l}{Basic jet properties} &
\multicolumn{3}{l}{What we observe} \\
\mr
Jet Plasma Bulk Velocity & $\beta$ & $v/c$$^\mathrm{(a)}$ &
Apparent Speed & $\beta_\mathrm{app}$ & $\beta \sin \theta / (1 - \beta \cos \theta)$$^\mathrm{(b)}$ \\
Intrinsic Luminosity & \multicolumn{2}{c}{$L_\mathrm{int}$} &
Luminosity & $L_\mathrm{obs}$ & $L_\mathrm{int} \times \delta^n$$^\mathrm{(c)}$ \\
Intrinsic Brightness Temperature & \multicolumn{2}{c}{$T_\mathrm{b,int}$}  &
Brightness Temperature & $T_\mathrm{b,obs}$ & $T_\mathrm{b,int}\times \delta$ \\
Bulk Lorentz Factor & $\Gamma$ & $(1-\beta^2)^{-1/2}$ &
Doppler Beaming & $\delta$ & $\Gamma^{-1}(1-\beta\cos\theta)^{-1}$ \\
\br
\multicolumn{6}{l}{\small $^{\rm a}$ $v$ is the speed of the plasma, and $c$ is the light speed.} \\
\multicolumn{6}{l}{\small $^{\rm b}$ $\theta$ is the viewing angle, notice that $\beta_\mathrm{app,max}\approx \Gamma$ when $\theta = \Gamma^{-1}$.} \\
\multicolumn{6}{l}{\small $^{\rm c}$ $n$ depends on the jet geometry and is typically in the range between 2 and 3.} \\
\end{tabular}
}
\]
\end{center}
\end{table}

Prior to the VLBA construction in the early 1990s, monitoring programs of 
AGN were difficult due to the sparse number of available antennas, the 
limited resources in terms of telescope time and recording media, and the 
difficulties to schedule repeated, homogeneous observations.  In 1994 we 
initiated a VLBA programme at 15\,GHz:  
the 2\,cm VLBA 
Survey\footnote{See \texttt{http://www.nrao.edu/2cmsurvey/}.} (see Table 
\ref{table:surveys}). 
The aim of the programme is to study the parsec-scale structure and 
kinematics of relativistic AGN jets to understand the acceleration and 
collimation of relativistic jets.
The program was continued as 
MOJAVE\footnote{\textbf{M}onitoring \textbf{O}f \textbf{J}ets in 
\textbf{A}ctive galactic nuclei with \textbf{V}LBA \textbf{E}xperiments, 
see \texttt{http://www.physics.purdue.edu/MOJAVE/}. } since 2003.  This 
effort, limited to declinations over $-20^\circ$, is complemented since 2007 
by the TANAMI\footnote{\textbf{T}racking \textbf{A}ctive Galactic 
\textbf{N}uclei with the \textbf{A}ustralian South-African 
\textbf{M}illiarcsecond \textbf{I}nterferometry, 
\texttt{http://http://pulsar.sternwarte.uni-erlangen.de/tanami/}.} 
program for the Southern skies ($\delta<-30^\circ$).  
The nominal resolutions of the images are slightly below 
1\,mas, and with typical overall on-source observing times of 1\,hr, dynamic 
ranges up to 1000:1 are achieved.  To monitor the jets in the radio 
sources, the feature positions and sizes of the radio images in the sky 
are measured directly, or alternatively, models are applied to the 
interferometric visibility data.  After several observations, their time 
evolution is studied.

Our effort in this field is complementary with further kinematic surveys 
performed at longer wavelengths (e.g., at $\lambda$3.6\,cm with a global 
array by Piner \textit{et al} (2007) and at $\lambda$6\,cm on the Caltech-Jodrell 
Flat Spectrum sample by and Britzen \textit{et al} 2008 
and references therein) 
and shorter ones ($\lambda$7\,mm monitoring on a smaller sample by Jorstad 
\textit{et al} (2001; 2005)).

\begin{table}
\caption{\label{table:surveys} Our VLBI Monitoring Programs}
\begin{center}
\[
\resizebox{\textwidth}{!}{%
\begin{tabular}{lllp{12cm}}
\br
Name & Time & Refs. & Description \\
\mr
2\,cm Survey & 1994--2002 & 1,2,3,4 & Over 200 sources imaged regularly with the VLBA at 15\,GHz \\
MOJAVE-I & 2002--2006 & 5,6,7,8 & Full linear polarisation added, source list revised to include a flux-limited sample of 135 sources.  The selection criteria are: $\delta>-20^\circ$, $|b|>2.5^\circ$, $S_\mathrm{VLBA,\,15\,GHz,\,1994.0-2003.0}>1.5$\,Jy (2\,Jy for $\delta<0^\circ$).  \\
MOJAVE-II & 2006--2007 & 9 & Expanded to 192 jets (58 EGRET blazars with $\delta>-20^\circ$, 33 low-luminosity AGN ($<10^{26}\,$W\,Hz$^{-1}$ at 15\,GHz), and 11 jets from the 2\,cm Survey with unusual kinematics, including a single epoch on every source at 8.1/8.4/12.1/15.3\,GHz during 2006 \\
MOJAVE-III & 2008-- & 9 & It will include up to 100 additional \textit{FGST}-detected sources (see text). \\
TANAMI & 2007-- & 10 & Monitoring of 40 sources at the Southern hemisphere at 0.3 (0.9)-mas resolution at $\lambda$\,1.1~(3.6)\,cm.  To be expanded to $\sim$120 sources in 2008--09.  Sources below $\delta = -30^\circ$, selected based on EGRET and on radio flux density and luminosity. \\
\br
\multicolumn{4}{l}{\small 1: Kellermann \textit{et al} (1994); 2: Zensus \textit{et al} (2002), 3: Kellermann \textit{et al} (2004);  4: Kovalev \textit{et al} (2005)} \\
\multicolumn{4}{l}{\small 5: Lister and Homan (2005), 6: Homan and Lister (2006), 7: Cooper \textit{et al} (2007), 8: Cara and Lister (2008)} \\
\multicolumn{4}{l}{\small 9: M L Lister \textit{et al} (in prep.), 10: Kadler \textit{et al} (2007)} \\
\end{tabular}
}
\]
\end{center}
\end{table}

\section{Observations and selected results}

\subsection{Overall findings}

By collecting subsequent observations of the quasars, the changing 
positions of the jet features are measured, yielding apparent speed values 
far beyond 10--15\,$c$.
The same measurements in radio galaxies 
and BL\,Lac objects show usually lower values.  Detailed results for all 
the sources in the surveys are presented in Kellermann \textit{et al} (2004), 
E Ros \textit{et al} (in prep.) and M L Lister \textit{et al} (in prep.). 
Additional images, movies, and
preliminary results can
also be found on the MOJAVE web site.

\subsection{Selected images and jet motions}

\begin{figure}[t!]
\includegraphics[clip,height=0.40\textwidth]{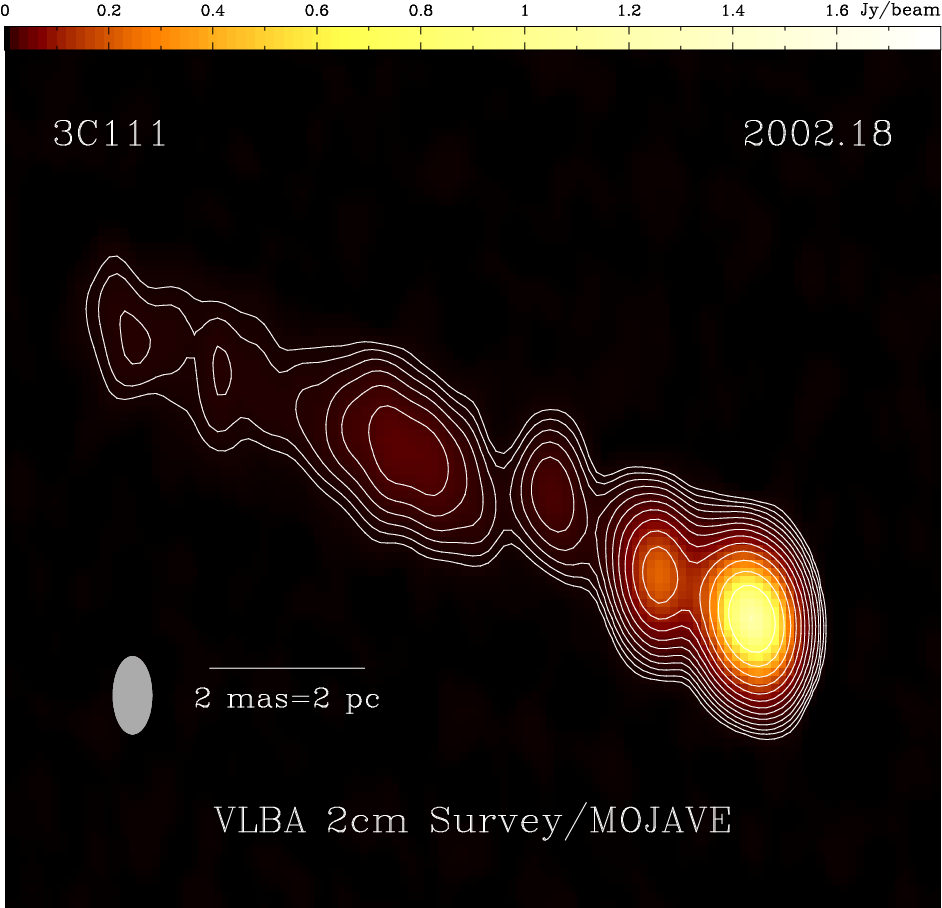}
\hfill
\includegraphics[clip,height=0.40\textwidth]{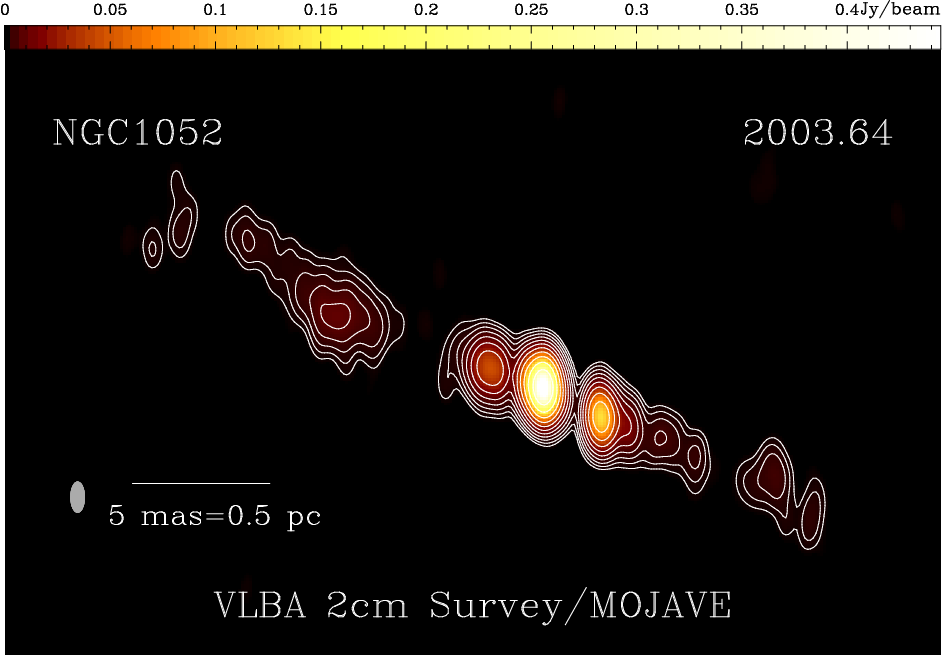}
%
%
\caption{\label{fig:images1} VLBA images of selected sources
from the 2\,cm/MOJAVE observing programme.
The \textit{left} panel shows 
an image of 
the Seyfert Galaxy 3C\,111 (4C\,+37.12, B0415+379, J0418+3801, with
a redshift $z$=0.0485), where trailing features after the ejection
of a major one have been reported;
see Kadler \textit{et al} 2008).
The \textit{right} panel shows an image of the twin
jet in the nearby Seyfert 2 galaxy NGC\,1052 (B0238$-$084, J0241$-$0815,
$z$=0.004930).  More details on this source are given in 
Ros \& Kadler (these proceedings).
}
\end{figure}

\begin{figure}[thbp]
%
\includegraphics[clip,height=0.47\textwidth]{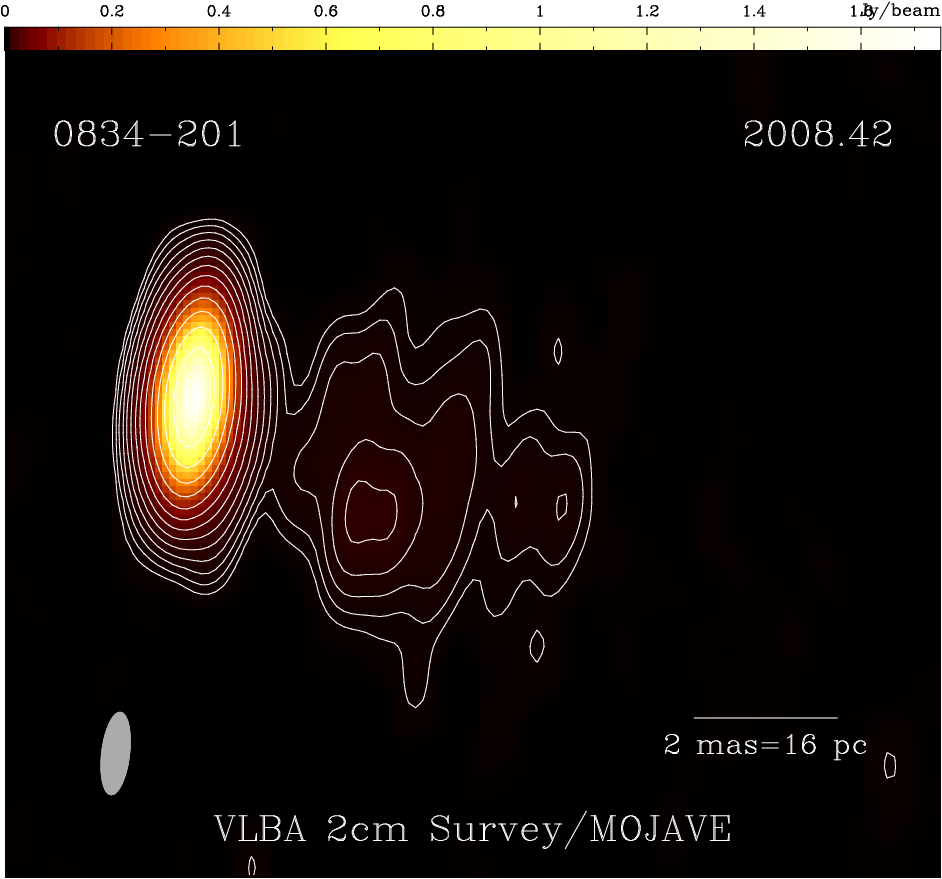}
\hfill
\includegraphics[clip,height=0.47\textwidth]{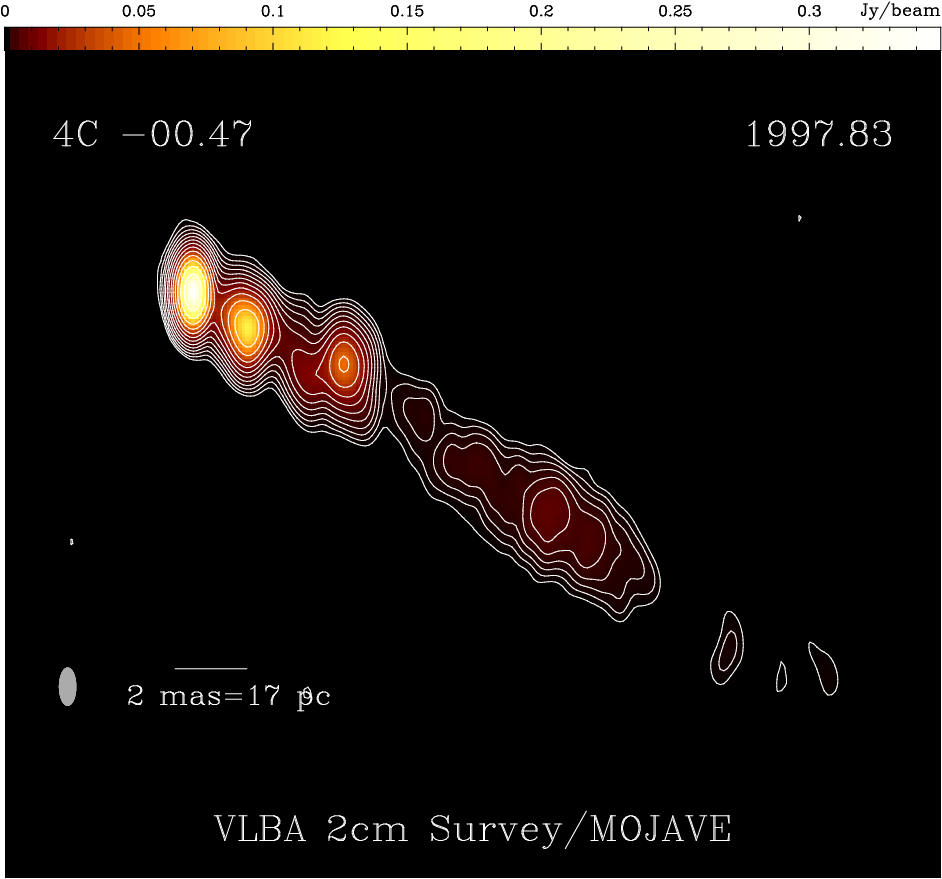}
%
%
\caption{\label{fig:images2} VLBA images of selected sources
from the 2\,cm/MOJAVE observing programme.
The \textit{left} panel shows the high-redshift quasar J0836$-$2016.
In contrast with the 8.4\,GHz image published in Ojha \textit{et al}
(2004), this source shows emission to the West in a core-jet
morphology.  Proper motions of the main features are shown
in Fig.\ \ref{fig:0834}. 
The \textit{right} panel shows the quasar 4C\,$-$00.47, where a more prominent
jet towards the south west.  Our monitoring
program do not show relevant superluminal motions in its components
(see Fig.\ \ref{fig:1148}).
}
\end{figure}

\begin{figure}[thbp]
%
\includegraphics[clip,height=0.46\textwidth]{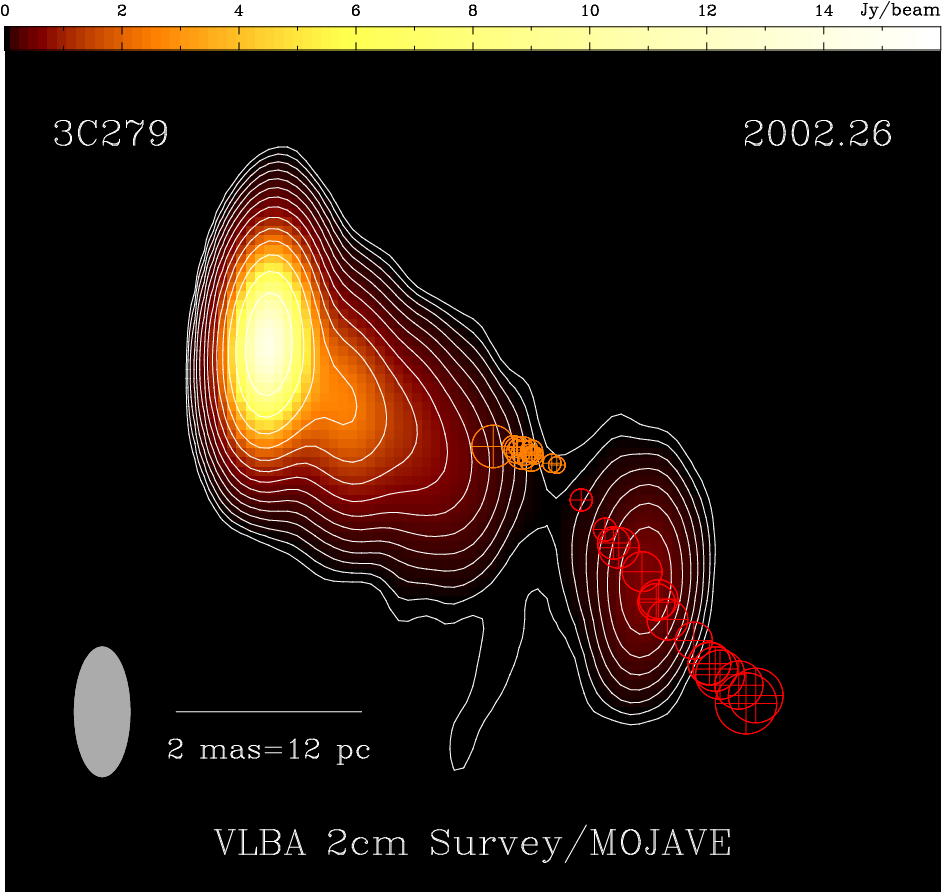}
\hfill
\includegraphics[clip,height=0.46\textwidth]{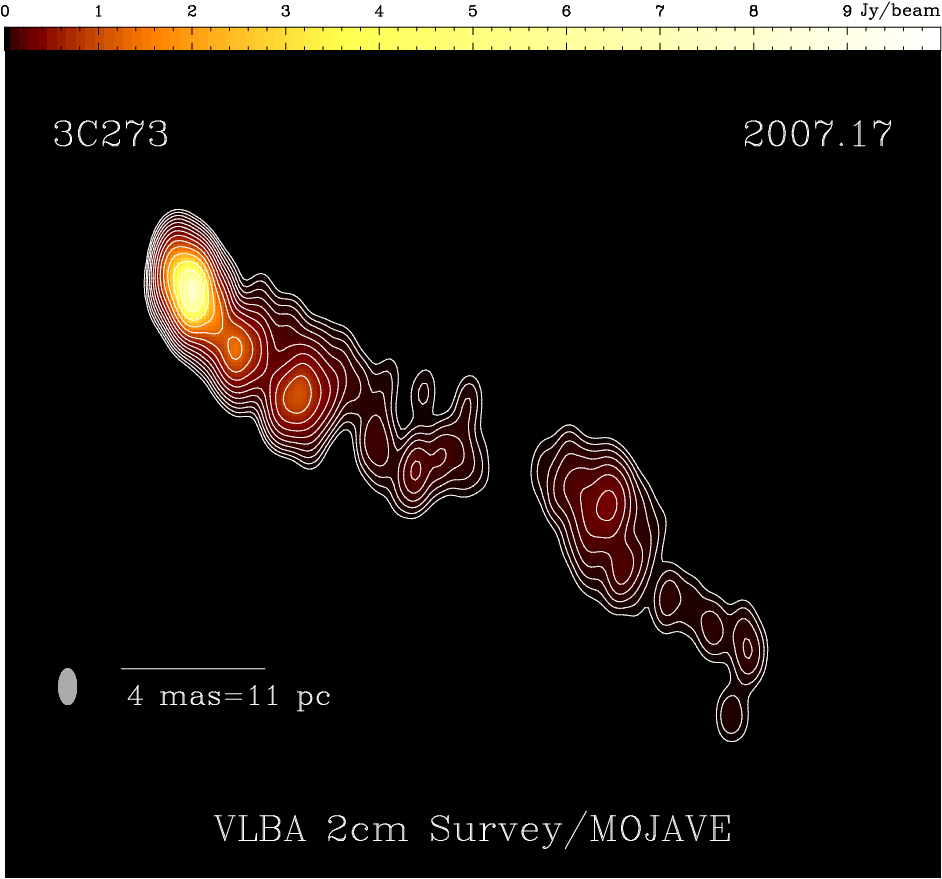}
\caption{\label{fig:images3} VLBA images of selected sources
from the 2\,cm/MOJAVE observing programme.
The \textit{left} panel shows the jet in 3C\,279 
(4C\,$-$05.55, B1253$-$4055, J1256$-$047, with
a redshift $z$=0.536).  
Homan \textit{et al} (2003) reported on the realignment on kiloparsec (kpc)
scales of the jet in 3C\,279 
as observed in milliarcsecond scales 
(de-projected kpc distance at a very small
viewing angle)
by a change in direction
and speed of a prominent feature.
The overlapped red, circled crosses show the positions for this
component.
The \textit{right} panel shows the jet in the well-known QSO 3C\,273
(B1226+023, J1229+0203, 4C\,+02.32, with 
a redshift of $z$=0.158).  
}
\end{figure}

In Figs.\ \ref{fig:images1}--\ref{fig:images3} we show selected contour images of some sources of our programme.  
Those are chosen because of their prominent jets and very complex structure.
The quasars 3C\,279 and 3C\,273 have being sampled very intensively 
in our programme,
and for our analysis we also could include a big amount of archival
data, since they are used regularly as calibrators and fringe finders
at the VLBA.  The nearby radio galaxies 3C\,111 
and NGC\,1052 are two cases of sources where we
have performed a detailed individual source analysis.  In the first
one we found trailing components after the ejection of a major feature
in the jet (Kadler \textit{et al} 2008), and for the second,
a twin jet is seen with mildly relativistic speed of 0.26$c$ traveling
downstream in both sides (see Ros \& Kadler, these proceedings, and
references therein).
The detailed kinematic results of these sources will be presented
in M L Lister \textit{et al} (in preparation).  We show in more
detail results for two sources not belonging to the complete MOJAVE-I
sample, for which we have measured kinematics.  These sources
have high redshifts and the measured motions are slow (not
at the abovementioned superluminal ranges over 10\,$c$).

\subsubsection{J0836$-$2016}
\begin{figure}[tbh]
\includegraphics[clip,width=0.65\textwidth]{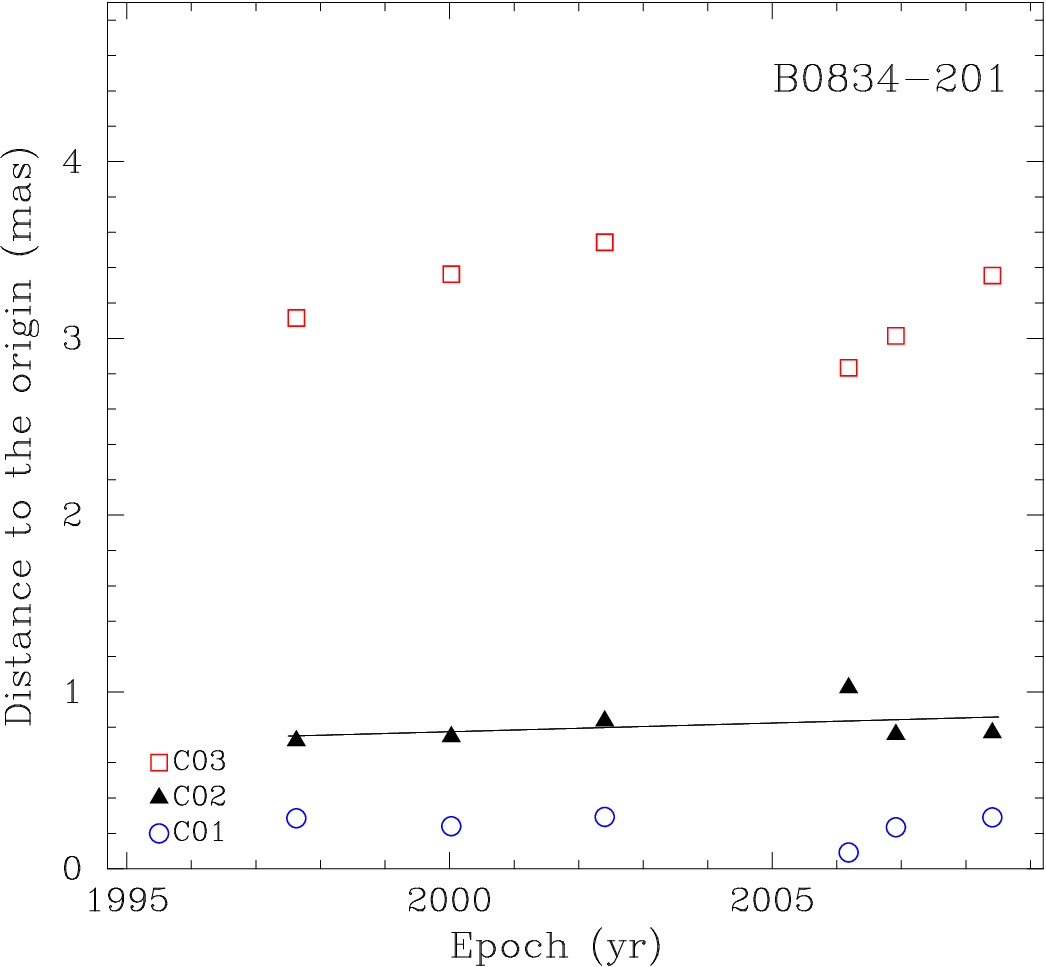}
\hfill
\begin{minipage}[b]{12pc}
\caption{\label{fig:0834} Proper motions in the jet of
B0834$-$201.  The points represent the distance of the
model fitted features with respect to the base of the jet
(core).  A linear fit to the distance of component
C02 is shown.  This component can be taken as stationary, since the fit
gives a proper motion of 10$\pm$12\,$\mu$as\,yr$^{-1}$ 
(corresponding to $\beta_\mathrm{app}$=1$\pm$1).
}
\end{minipage}
\end{figure}

This high redshift ($z$=2.752) QSO (OJ\,$-$257.5, B0834$-$201)
shows a compact core-jet structure (see Fig.\ \ref{fig:images2}) with
no big changes between epochs.  Preliminary kinematic
results are shown in Fig.\ \ref{fig:0834}.

\subsubsection{4C\,$-$00.47}
\begin{figure}[t!]
\includegraphics[clip,width=0.65\textwidth]{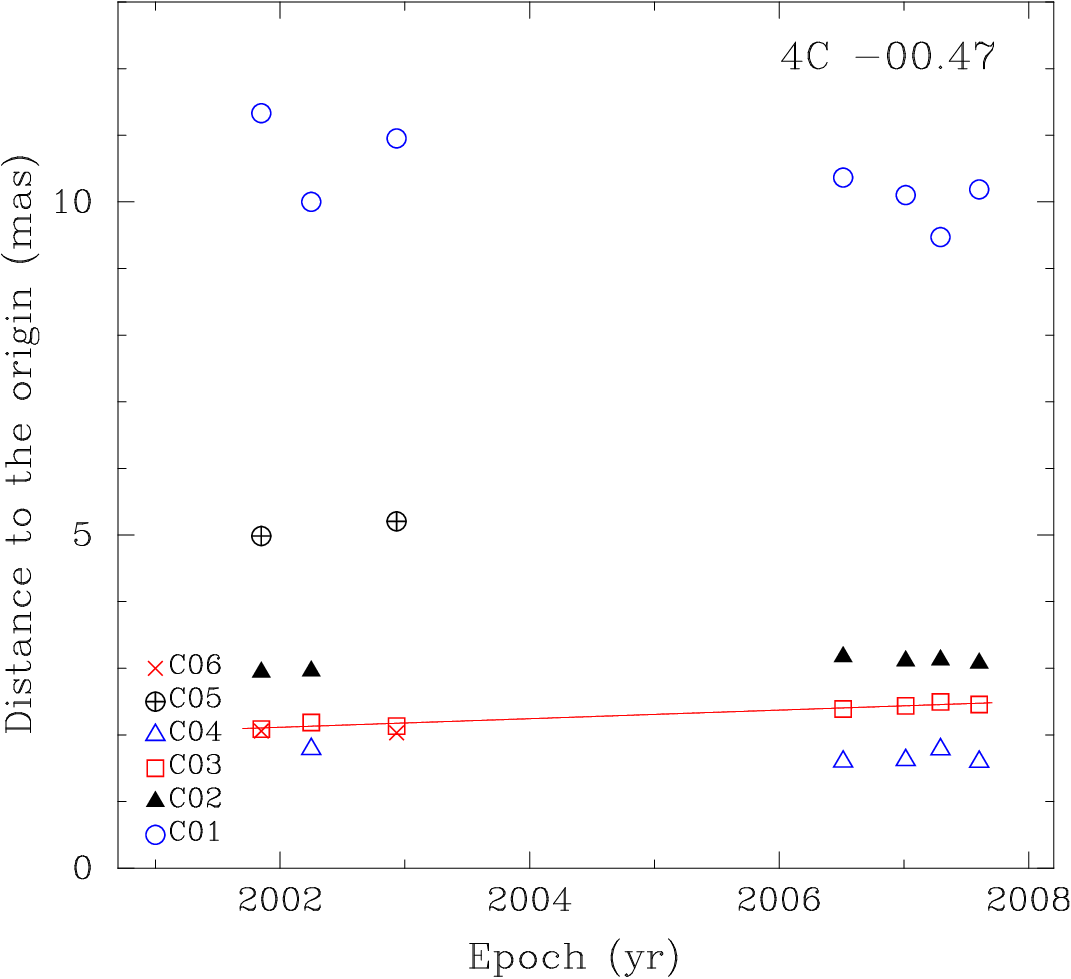}
\hfill
\begin{minipage}[b]{12pc}
\caption{\label{fig:1148} Proper motions in the jet of the
quasar 4C\,$-$00.47.  The relative distance to the innermost feature
(core) is shown as a function of time.  We have fitted a linear
speed to the positions of feature C03, which provides a sky motion
of 64$\pm$6\,$\mu$as\,yr$^{-1}$, which corresponds to 
$\beta_\mathrm{app}$=5.2$\pm$0.5 and a nominal ejection epoch
value of year 1969$\pm$3.
}
\end{minipage}
\end{figure}

This QSO (OM\,$-$080, B1148$-$001, J1150$-$0024, at a
redshift of $z$=1.975, presents an extended jet to the Southwest 
(see Fig.\ \ref{fig:images2}).  Its imaging is relatively challenging
due to its proximity to the equator, what makes a
$(u,v)$-coverage dominated by non-crossing east-west traces.  However,
our images yield the quality to trace components among different 
observing epochs.
Its jet does not show 
high speeds, and the kinematical results are compatible with
stationary components (see Fig.\ \ref{fig:1148}).

\section{Kinematics and \textit{FGST}}

The \textit{Fermi Gamma-ray Space Telescope} (\textit{FGST}, earlier known as 
\textit{GLAST}), launched in June 2008, is going to provide an 
all-sky $\gamma$-ray coverage every few hours.
Following the discoveries
of EGRET, the gamma-ray-sky will be dominated by relativistically-beamed 
AGN (blazars).  Since the bulk of $\gamma$-ray emission is expected
to be produced close to the jet base, our high-resolution
observations are crucial to understand the mechanisms in the
jet production and the overall physical properties
of the jets, complementing the \textit{FGST} results.
The first data of \textit{FGST} are being processed, and new
avenues are open for the multi-band studies of AGN by combining
VLBI, \textit{FGST} in the $\gamma$-ray regime, and \textit{Swift} at
intermediate bands in coordinated observations.

\ack
This paper is based on observations made with the VLBA, which is a 
facility
of the National Radio Astronomy Observatory which is operated by 
Associated Universities Inc., under a cooperative agreement with 
the USA National Science Foundation.  The MOJAVE project is supported 
under the USA National Science Foundation grant AST-0406923 and
NASA-\textit{Fermi} grant NNX08AV67G. 
We are very thankful to the 2\,cm~Survey/MOJAVE and the TANAMI teams.
C.\ M.\ Fromm produced the color images of the radio sources.

\section*{References}
\begin{thereferences} 


\item Britzen S, Vermeulen R C, Campbell R M, Taylor, G B, Pearson T J, Readhead A C S, Xu W, Browne I W, Henstock D R and Wilkinson, P 2008 \textit{A\&A} \textbf{484} 119--142

\item Cara M and Lister M L 2008 \textit{AJ} \textbf{674} 111-121 


\item Cooper N J, Lister M L and Kochanczyk M D 2007 \textit{ApJS} \textbf{171} 376--388 

\item Homan D C, Lister M L, Kellermann K I, Cohen M H, Ros E, Zensus J A, Kadler M and Vermeulen R C 2003 \textit{ApJ} \textbf{589} L9--L12 

\item Homan D C and Lister M L 2006 \textit{AJ} \textbf {131} 1262--1279 

\item Jorstad S G, Marscher A P, Mattox J R, Wehrle A E, Bloom S D and Yurchenko A V 2001 \textit{ApJSS} \textbf{134} 181--240 %

\item Jorstad S G \textit{et al} 2005 \textbf{AJ} \textbf{130} 1418--1465

\item Kadler M, Ojha R, Tingay S, Lovell J and TANAMI Collaboration 2007 \textit{Bull Am Astron Soc} \textbf{38} 732

\item Kadler M \textit{et al} 
2008 \textit{ApJ} \textbf{680} 867--884

\item Kellermann K I, Vermeulen R C, Zensus J A and Cohen M H 1998 \textit{AJ} \textbf{115} 1295--1318

\item Kellermann K I \textit{et al}
2007 \textit{Ap\&SS} \textbf{311} 231

\item Kellermann K I, Lister M L, Homan D C, Vermeulen R C, Cohen M H, Ros E, Kadler M, Zensus J A and Kovalev Y Y 2004 \textit{ApJ} \textbf{609} 539--63



\item Kovalev Y Y \textit{et al} 2005 \textit{AJ} \textbf{130} 2473--2505 

\item Kovalev Y Y, Lister M L, Homan D C and Kellermann K I 2007 \textit{ApJ} \textbf{668} L27--L30

\item Lister M L and Homan D C 2005 \textit{AJ} \textbf{130} pp 1389--1417 

\item Ojha R \textit{et al} 2004 \textit{AJ} \textbf{127} 3609--3621 

\item Piner B G, Mahmud M, Fey A L and Gospodinova K 2007 \textit{ApJ} \textbf{133} 2357-- 

\item Zensus J A, Ros E, Kellermann K I, Cohen M H, Vermeulen R C and Kadler M 2002 \textit{AJ} \textbf{124} 662--674
\end{thereferences}

\end{document}